\documentclass[10pt,aps,prl,twocolumn,showpacs,preprintnumbers,amsmath,amssymb]{revtex4-1}

\usepackage{graphicx}
\usepackage{bm}
\usepackage[caption=false]{subfig}
\usepackage{amssymb,amsmath}
\usepackage{verbatim}
\usepackage{color}
\begin{document}

\preprint{IQSL-Grating Cavity}

\title{Strong Optical Confinement between Non-periodic Flat Dielectric Gratings}
\author{Jingjing Li}
\email{jingjingl@hp.com, jingjingli2004@gmail.com}
\affiliation{Hewlett-Packard Laboratories, 1501 Page Mill Road,
Palo Alto, CA 94304--1123}

\author{David Fattal}
\affiliation{Hewlett-Packard Laboratories, 1501 Page Mill Road,
Palo Alto, CA 94304--1123}

\author{Marco Fiorentino}
\affiliation{Hewlett-Packard Laboratories, 1501 Page Mill Road,
Palo Alto, CA 94304--1123}

\author{Raymond G.\ Beausoleil}
\affiliation{Hewlett-Packard Laboratories, 1501 Page Mill Road,
Palo Alto, CA 94304--1123}

\date{\today}

\begin{abstract}
We present a novel design of optical micro-cavity where the optical energy resides primarily in free space, therefore is readily accessible to foreign objects such as atoms, molecules, mechanical resonators, etc.  We describe the physics of these resonators, and propose a design method based on stochastic optimization.  Cavity designs with diffraction-limited mode volumes and quality factors in the range of $10^4$--$10^6$ are presented.  With a purely planar geometry, the cavity can be easily integrated on-chip using conventional micro- and nano- fabrication processes.
\end{abstract}

\pacs{42.60.Da, 42.15.Dp, 42.25.Fx, 87.80.Cc}

\maketitle
Optical cavity has become an indispensable element in a wide range of applications such as lasers, sensors, cavity quantum electrodynamics (CQED), optomechanics, optical traping and manipulation, etc.\cite{Vahala03_Nature, Vuckovic06_NatPhys, Vollmer08_NatMet, Kimble11_NatPhys, Vuckovic10_PRL, Kippenberg09_NatPhys, Dienerowitz08_JN}.  For these applications, a high quality factor ($Q$) and/or a small effective mode volume (defined as $V=\int\epsilon|E|^2 \mathrm{dv}/\left(\epsilon|E|^2/2\right)_\mathrm{max}$)\cite{Painter_Science_99} is usually important, with the first a measure of the photon temporal confinement, and the second inversely proportional to the electric field per confined photon.  The ratio of the two ($Q/V$), which is proportional to the Purcell factor\cite{Purcell_PR_46}, represents the increase of the spontaneous emission rate for a light emitter and is another figure of merit in cavity design.  It plays an important role in determining the lifetime of an excited atom or a quantum dot contained within the cavity.  Various on-chip optical cavities of micrometer size have been designed and fabricated that can be categorized into three types, depending on their light-confinement mechanisms: Fabry-P\'erot (FP) cavities, cavities with whispering-gallery (WG) modes (micro-rings, disks, spheres, toroids, etc.) and two dimensional (2D) photonic crystal (PhC) defect cavities\cite{Vahala03_Nature}.  Integrated FP cavities---such as those used in Vertical Cavity Surface Emission Lasers (VCSELs)---go to resonance when light interferes constructively after a round trip between two mirrors that are usually distributed Bragg reflectors (DBRs).  High $Q$-s close to $10^6$ are reported, but a current-confining oxide aperture or a waveguide structure such as a micropillar is usually required for lateral mode confinement\cite{Stoltz05_APL, Forchel07_APL}, while the mode often penetrates into the DBRs for a depth up to several wavelengths. WG cavities confine the light along a circular or spherical wall through total internal reflection (TIR), and build up a resonant mode when the phase delay in a full circular trip is the multiple of $2\pi$.  Very high $Q$ values (up to $10^9$) were achieved for devices of large size ($\sim 100\mu\mathrm{m}$) with less ideal $V$\cite{Vahala03_Nature}.  Micro-rings of small radius were also reported with the trade off of lower $Q$-s due to a larger bending loss\cite{Xu08_OE}.  2D PhC defect cavities confine the light through a photonic bandgap in the lateral directions and TIR in the vertical.  They are capable of achieving modes with small mode volumes ($\sim \left (\lambda_0/n\right )^3$) but usually lower $Q$s compared to the best of WG cavities, mainly due to the difficulty in confining the mode in the vertical direction when the size is small \cite{Vuckovic06_NatPhys, Vuckovic10_PRL, Notomi_APL_2007}.
\begin{figure}
\begin{center}
\subfloat[][]{
	\label{subfig:Parabolic:Mirror}
	\includegraphics{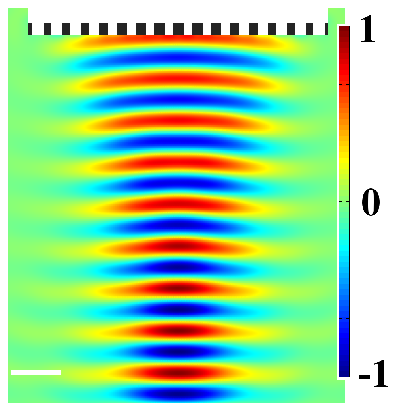}
}
\subfloat[][]{
	\label{subfig:Parabolic:Cavity}
	\includegraphics{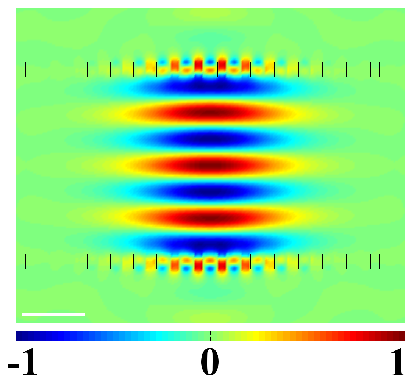}
}
\end{center}
\caption{\label{Fig:Parabolic} (a) Plot of the instantaneous distribution of $E_x$ (horizontal in the plane of the page) reflected by a parabolic grating mirror, normalized to its maximum value.  The incident field has a flat curvature and an operating wavelength $\lambda_0 = 835$~nm. (b) Plot of $E_x$ at 835~nm  for a cavity composed of two such mirrors, normalized to its maximum value. The white scale bars have a length of 1~$\mu$m.}
\end{figure}

One essential issue in many applications using optical cavities is the accessibility for particles to regions of high electric field.  This determines the coupling strength between the light emitter and the cavity in CQED, the sensitivity of an optical-cavity based sensor, and the strength of the force when the cavity mode is used to influence nano-mechanical structures or to trap and manipulate nano particles.  While each of the cavity designs discussed above manage to achieve high $Q$ and/or small $V$, in reality they all show disadvantage in this criterion.  For WG and PhC cavities that utilize TIR, the high-field regions are either contained inside the dielectric or in a small region close to the dielectric sidewalls, making the access to this region very difficult.  In reality, the atoms, molecules or nanomechanical resonators are either brought into the evanescent tail of the mode outside of the cavity\cite{Vollmer08_NatMet, Kimble11_NatPhys, Kippenberg09_NatPhys}, or planted inside the dielectric body during the fabrication\cite{Vuckovic10_PRL}.  Although an FP cavity without any medium in between the two mirrors could in principle provide a mode accessible to foreign particles, the lack of lateral confinement in this case would inevitably result in a largely extended, or even instable, resonant mode\cite{Yariv75_Book}.  

In this paper we present a new type of optical cavity with a reasonably high $Q$, a diffraction-limited mode volume, and much-improved field accessibility.  To introduce our cavity design, let us first examine a conventional FP cavity, but consists of two identical dielectric mirrors working as parabolic reflectors.  Following an approach we recently developed in Ref.~\onlinecite{Fattal10_NatPhot}, these dielectric mirrors were designed by introducing a smooth variation of the duty cycle (DC, the ratio of the width of the dielectric groove to the period of the grating) of a dielectric grating.  By locally tuning the DC, the reflection phase of the mirror is tailored to be a parabolic distribution to the position while the reflection magnitude maintains high everywhere, so that the mirror works as a parabolic reflector while remaining a planar geometry\cite{Fattal10_NatPhot}. Fig.~\ref{subfig:Parabolic:Mirror} shows a 2D finite difference time domain (FDTD) simulation of the behavior of such a mirror under a Gaussian beam incidence.  This simulation (and others below) was done using an open source package MEEP\cite{ref:oskooi2010mff}.  The mirror has a focus length of $13\mu\mathrm{m}$ and a reflectance of $\simeq98\%$.  The resonant mode of a cavity composed of two of such mirrors is shown in Fig.~\ref{subfig:Parabolic:Cavity} together with the outline of the geometry.  Notice that, the mode has a $Q=2,000$ with a full width at half maximum (FWHM) of $1.5\lambda_0$.  These can not be achieved for VCSEL cavities of planar reflectors without a mode-confining mechanism.

Compared to conventional VCSEL cavities with DBRs, the one shown in Fig.~\ref{subfig:Parabolic:Cavity} supports a strictly stable mode of standing Gaussian beam with a finite cross section, with a $\sim 1/4\lambda_0$ field penetration into the dielectric mirrors.  It is in principle similar to a conventional FP resonator with curved mirrors but is compatible to a planar fabrication process.  The relevant parameters such as the resonance frequency, the waist, and the quality factor can be predicted from the effective curvature of the mirrors and their separation using formulas for FP with curved mirrors\cite{Yariv75_Book}.  However, this approach would not be successful when we try to further reduce the mode cross-section to sub-wavelength size.  A Gaussian beam with a cross section even smaller would be too diverging to satisfy the longitudinally slow-varying condition, thus the underlying physics of the cavity shown in Fig.~\ref{subfig:Parabolic:Cavity} no longer holds.  Also, the optical mode may only interact with a few individual dielectric ``grooves'' (or ``posts''), whose collective behavior as a grating is no longer relevant.  Rather, the dielectric grooves would behave like independent, near-resonant scatterers, and the global behavior of the assembly as a cavity is the result of the combined effect of these coupled scatterers.  If the structure could be designed so that the radiation from the scatterers interferes destructively in the far-field, a high-Q optical mode can
be obtained.   
\begin{figure}
\begin{center}
\subfloat[][]{
	\label{subfig:2DCavity:Schem}
	\includegraphics[width=1.5in]{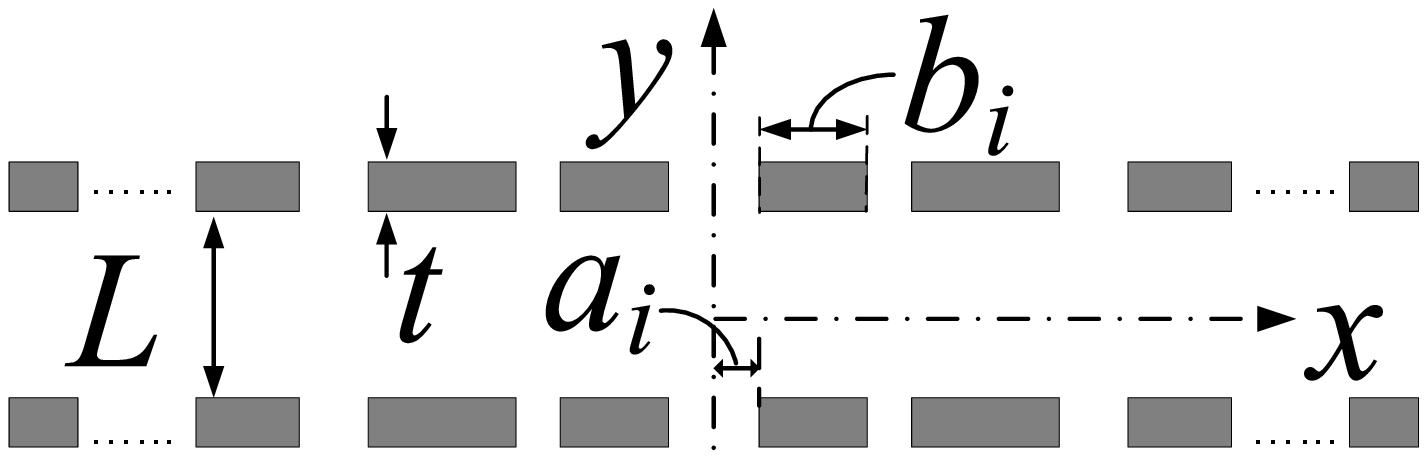}
}
\subfloat[][]{
	\label{subfig:2DCavity:ESq}
	\includegraphics{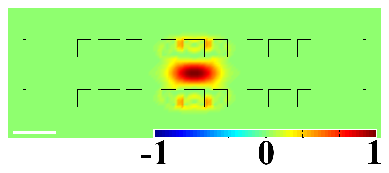}
}\\
\subfloat[][]{
	\label{subfig:2DCavity:Hz}
	\includegraphics{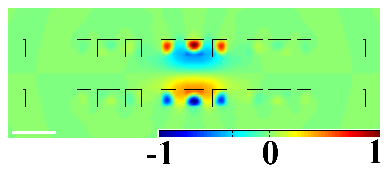}
}
\subfloat[][]{
	\label{subfig:2DCavity:Ex}
	\includegraphics{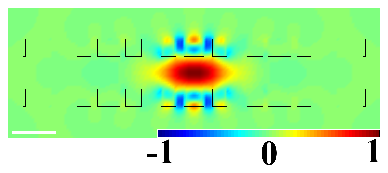}
}\\
\subfloat[][]{
\label{fig:ESqDistribution}
\includegraphics{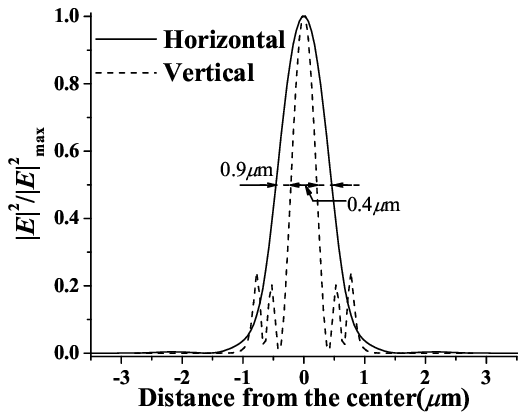}
}
\end{center}
\caption{\label{fig:2DCavity}(a) The schematic of the cavity design. (b)-(d) The instantaneous field distributions of one cavity design, normalized to their respective maxima: (b) $|E|^2$; (c) $H_z$; (d) $E_x$.  The white scale bars in these plots have a length of 1~$\mu$m. (e) $|E|^2$ distribution along the horizontal and vertical symmetry axis.}
\end{figure}

A deterministic design of the resonance features of the grooves and their mutual coupling would be extremely complicated.  As a first attempt to test the feasibility of the idea we use a stochastic optimization method to search for a cavity design with high $Q$ and/or small $V$.  We first try to design an optical cavity with a structure symmetric with respect to the $x$ and $y$ axes (Fig.~\ref{subfig:2DCavity:Schem}), and we look for a fundamental TM mode near $\lambda_0=1.55\mu\text{m}$ with an asymmetric transverse magnetic field distribution along the $x$ axis.  In each quadrant, we use $N$ dielectric grooves of refractive index $n=3.48$ (silicon, around $\lambda_0=1.55\mu\text{m}$), where $N$ is determined before hand.  The width and position of each groove is determined by parameters $a_i$ and $b_i$ (as shown in Fig.\ref{subfig:2DCavity:Schem}). Some or all of the $a_i$ and $b_i$, together with the groove thickness $t$ and cavity length $L$ are used as optimization parameters.  This corresponds to $14$ degrees of freedom for $N = 6$.  An exhaustive search is not practical, and a stochastic optimization has proved to be essential here.  A parallel genetic algorithm (GA) is used in our study, and we refer the readers to Ref.~\onlinecite{Rahmat_Samii_IEEEAPM_97} for a  brief, yet informative introduction to GA and its applications to EM problems.  In a GA process, each optimization parameter is binary-coded and the codes are then concatenated to form a binary string (a ``chromosome'').  Each chromosome represents a possible cavity design whose behavior is simulated by an EM solver and its ``fitness'', a quantitative description of how well the design approaches the desired performance, is calculated.  The GA evolves towards a design with maximum fitness.  Different choices of the fitness function give preference to different aspects of performance, thus resulting in different cavity designs.  In our study we use either $Q$ or $Q/V$ as the fitness function.  The EM simulation is done on MEEP employing its parallel algorithm.  The optimization process is carried out on a computer cluster using up to $160$ cores, with multiple cavity simulations, each of which uses $4$ cores, running at the same time.  Because of this massively parallel strategy, the optimization process converges  quickly ($\sim$20--30 hours) in spite of the large search space.

The optimization process indeed yields cavities with properties as expected.  The normalized instantaneous distributions of $|E|^2$, $H_z$, and $E_x$ for one of the results are shown in Fig.~\ref{subfig:2DCavity:ESq}, \ref{subfig:2DCavity:Hz}, and \ref{subfig:2DCavity:Ex}, respectively.  In this case, the thickness of the grooves is $t = 400$~nm and the cavity length is $L = 759\text{nm}$.  The structure has a total width of about $8\mu\text{m}$ and supports a cavity mode with a free space wavelength $\lambda_0 = 1542.6\text{nm}$ with a quality factor $Q = 14,697$.  Notice that the profile of the grating-like structure is fairly ``irregular'' compared to that of the parabolic mirror shown in Fig.~\ref{Fig:Parabolic}.  The cavity does show reasonable tolerance of variations in the structural parameters around the optimized design.  For example, $Q$ decreases as $L$ deviates from the optimal value, but stays above $10,000$ for $750\text{nm}< L < 770\text{nm}$, when the resonant wavelength varies between $1539\text{nm}$ and $1548\text{nm}$.  The cavity mode, which otherwise would not exist, demonstrates the radiation cancellation provided by each dielectric groove, and the quality factor is determined by the residue radiation into the far field that is not ideally canceled by the grooves.

In addition to the fact that it can be built using planar technology, two other features of the cavity deserve specific mentioning.  First, the mode is concentrated around the center of the structure confined in free space between the dielectric layers.  As shown in Fig.\ref{fig:ESqDistribution}, along the $y$ axis the field is confined primarily between the two dielectric gratings with a full width at half maximum (FWHM) of only $\sim400~\text{nm}$.  Along the $x$ axis, where no mode-confining structure exists, the FWHM is $\sim$900~nm, well below the vacuum wavelength.  The equivalent volume for the 2D mode is $V=0.283\, \mu\mathrm{m}^2 = 0.119\, \lambda_0^2$.  Second, the radiation leaked into the far-zone forms multiple sharp beams, with the dominant beam pointing to $\pm y$ directions for many cases.  For the radiation, the dielectric grooves work as an antenna array of large aperture fed by the cavity mode of small cross section.  By reciprocity, we can also use this ``antenna array'' at the receiving mode, and to excite the cavity mode using a plane incident wave.  This is indeed proved by numerical simulation.  This type of excitation is much more convenient compared to the excitation of WG cavities where waveguides and phase-matching are required, or that of PhC cavities where tapered fibers are used.
\begin{figure}
\begin{center}
\subfloat[][]{
	\label{subfig:3Dcav:geom}
	\includegraphics[width=1.8in]{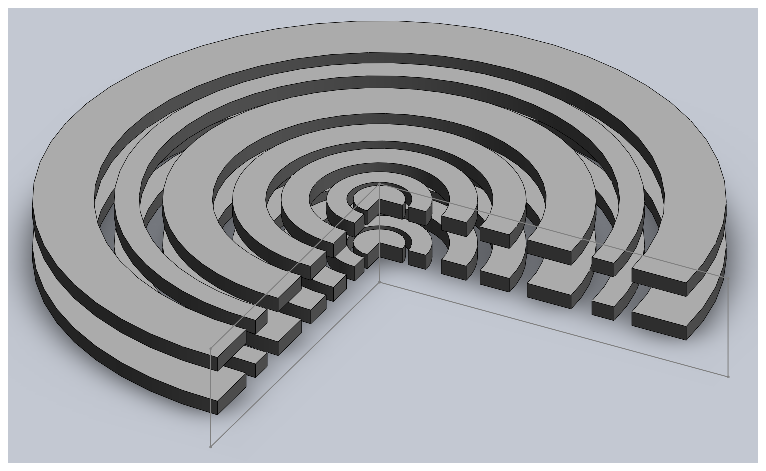}
	}\\
\subfloat[][]{
	\label{subfig:3Dcav:field}
	\includegraphics{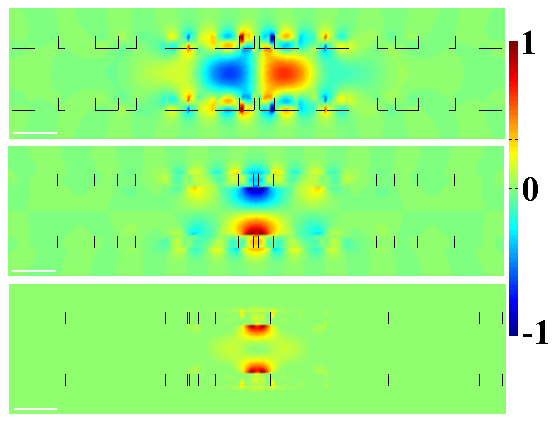}
	}
\caption{\label{fig:3DCavity} (a) A schematic of the 3D cylindrical cavity. (b) The instantaneous distribution of $E_r$ (top), $E_z$ (center) and $|E|^2$ (bottom) of the cavity mode, all normalized to their respective maxima. The white scale bars have a length of 1~$\mu$m.}
\end{center}
\end{figure}

The simulation cases shown so far are for planar, 2D structures invariant along one of the Cartesian axes.  We can also design 3D cavity that has a cylindrical symmetry.  In this case the cavity consists of two identical layers comprising multiple concentric rings of different widths, as schematically shown in Fig.~\ref{subfig:3Dcav:geom}, and we look for solutions in which the angular dependence of the fields is proportional to $e^{i m \phi}$, where $\phi$ is the azimuthal angle and $m$ is an integer.  Like the linear grooves, cylindrical grooves exhibit local resonances that can be harnessed to support localized 3D cavity modes.  Thanks to the rotational symmetry of the structure the FDTD simulation reduces to an effective 2D problem that can be quickly solved.  We apply the same GA optimization to the cylindrical problem using the thickness, widths and radius of the dielectric rings as optimization parameters.  In what follows we limit our analysis to TM modes with $m=0$ (i.e. the only non-zero component of magnetic field is $H_\phi$).  Cavities supporting $m\neq 0$ modes can also be designed using a similar approach.  Fig.~\ref{subfig:3Dcav:field} shows instantaneous distribution of $E_r$, $E_z$, and $|E|^2$ on the cross section through the symmetry axis for a typical design of $\lambda_0=1552.6$~nm and $Q=60,435$.  Similar to the 2D cavity case, the mode is largely confined in free space, around the symmetry axis in between the dielectric layers, with an effective mode volume of $V=0.637\, \mu\mathrm{m}^3 = 0.264\, \lambda_0^3$. The cavity can also be excited through an azimuthally polarized beam at normal incidence.
\begin{figure}
\includegraphics{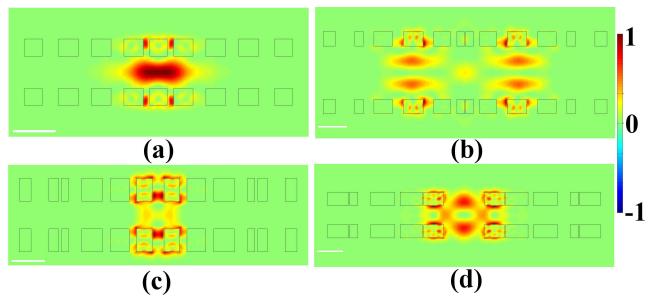}
\caption{\label{fig:MoreCavities}$|E|^2$ distributions (normalized to their respective maxima) for various cavity designs.  (a) 2D cavity, limited size, regular groove shape, $Q=78,260$.  (b) 3D cylindrical cavity, higher order mode, $Q=92,384$.  (c) 2D cavity, thick dielectric layer, $Q=601,320$. (d) 3D cylindrical cavity, open central part, $Q=14,337$. The white scale bars have a length of 1~$\mu$m.}
\end{figure}

The examples shown above are only a subset of the designs we have achieved.  With different choices of fitness function ($Q$, $1/V$, or $Q/V$), types and ranges of the optimization parameters (groove thickness, cavity length, some or all the sizes and positions of the grooves, etc.), cavities with widely varying characteristics can be designed.  A few other typical examples are shown in Fig.~\ref{fig:MoreCavities}, where Fig.~\ref{fig:MoreCavities}a and c show 2D designs while those in Fig.~\ref{fig:MoreCavities}b and d are cylindrical 3D cavities.  In the design of Fig.~\ref{fig:MoreCavities}a we limit the over all lateral size of the cavity to be smaller than 6.5~$\mu$m and use regularly shaped dielectric grooves with aspect ratio $<1$, to reduce the fabrication difficulty.  The mode is wider compared to that in Fig.\ref{fig:2DCavity} but the effective 2D mode volume is still smaller than $\lambda_0^2$ with $Q>78,000$.  For the design in Fig.~\ref{fig:MoreCavities}c, thicker dielectric grooves are used and more EM energy is stored inside the dielectric, resulting in a quality factor $Q>600,000$.  In Fig.~\ref{fig:MoreCavities}b a cavity of a higher order longitudinal mode with a quality factor $Q>90,000$ is shown. In the cavity in Fig.~\ref{fig:MoreCavities}d a large portion in the center of the structure is left open for convenient access of the mode field.  The optical field is still concentrated around the symmetry axis with $Q>14,000$.

We emphasize here that the function of the two layers of dielectric grooves in this cavity design is fundamentally different to that of the mirrors in a FP cavity.  A single layer of this structure does not show any appreciable reflectivity, and the field in between the two layers can not be described as a back-and-forth traveling wave.  As we see in the plots, a large portion of the resonant mode, including the maximum of the field, stands in free space, freely accessible by molecules, light emitters, nanomechanical resonators, etc.  This would provide a much stronger effect compared to the evanescent field coupling used in existing references.  The cavity holds a resonant mode with an open structure, essentially works as a ``cage'' for photons.  Because of these features, we believe it can be a unique candidate for a wide range of applications, many of which are currently under study.

%

\end{document}